\pdfoutput=1

\documentclass{pnastwo}

\usepackage{graphicx}

\usepackage{amssymb,amsfonts,amsmath,color}

 \newcommand{\etal}{\mbox{\emph{et al.\ }}}
 \newcommand{\bom}{{\mbox{\boldmath $\omega$}}}
 \newcommand{\bs}{{\bf {s}}}

\newcommand{\bv}{{\bf {v}}}

\contributor{Submitted to Proceedings of the National Academy of Sciences of the United States of America}
\copyrightyear{2013}
\issuedate{Issue Date}
\volume{Volume}
\issuenumber{Issue Number}
\url{www.pnas.org/cgi/doi/10.1073/pnas.1312535111} 
\url{} 
\footlineauthor{H\"anninen and Baggaley} 

\begin{document}



\title{Vortex filament method as a tool for computational visualization of quantum turbulence}





\author{
Risto H\"anninen\affil{1}{O.V. Lounasmaa Laboratory, Aalto University, FI-00076 Aalto, Finland}
\and 
Andrew W. Baggaley\affil{2}{School of Mathematics and Statistics, University of Glasgow, Glasgow, G12 8QW, UK}
}

\contributor{Submitted to Proceedings of the National Academy of Sciences of the United States of America}
\contributor{} 

\maketitle

\begin{article}

\begin{abstract} 
Vortex filament model has become a standard and powerful tool to visualize the motion of quantized vortices in helium superfluids. In this article, we present an overview of the method and highlight its impact in aiding our understanding of quantum turbulence, particularly superfluid helium. We present an analysis of the structure and arrangement of quantized vortices. Our results are in agreement with previous studies showing that under certain conditions, vortices form coherent bundles, which allows for classical vortex stretching, giving quantum turbulence a classical nature. We also offer an explanation for differences between the observed properties of counterflow and pure superflow turbulence in a pipe. Finally, we suggest a mechanism for the generation of coherent structures in the presence of normal fluid shear.
\end{abstract}







\dropcap{T}urbulence in fluid flows is universal, from galactic scales generated by supernova explosions 
down to an aggressively stirred cup of coffee. There is no debate that turbulence is important, and yet no 
satisfactory theory exists. Turbulence is built by rotational motions, typically over a wide range of scales, 
interacting and mediating a transfer of energy to scales at which it can be dissipated effectively. The 
motivation for K\"uchemann's famous quote ``vortices are the sinews and muscles of fluid motions'' is clear. 
If this is true, then quantum turbulence (QT) represents the skeleton of turbulence, and offers a method of 
attacking the turbulence problem in perhaps its simplest form.

QT is a tangle of discrete, thin vortex filaments, each carrying a fixed circulation. It is typically 
studied in cryogenically cooled helium \cite{DonnellyBook,carlo-book} and, more recently, in atomic 
Bose--Einstein condensates \cite{bagnato}. These substances are examples of so-called quantum fluids: 
fluids for which certain physical properties cannot be described classically but depend on quantum mechanics. 
The quantisation of vorticity is one marked difference between quantum and classical fluids. Another is their 
two-fluid nature; they consist of a viscous normal fluid component and an inviscid superfluid component 
coupled by a mutual friction. The relative densities of these components are temperature dependent. 

Despite these marked differences, it is now the consensus opinion that QT is capable of exhibiting many 
of the statistical properties of classical turbulence, including the famed Kolmogorov scaling \cite{VinenJLTP2002}. 
Hence, QT has the potential to offer new insights into vortex dynamics and the role they play in the 
dynamics of turbulence. In addition, QT offers many interesting problems in its own right. However, QT, 
more so than classical turbulence, suffers from poor visualization of the flow in experiments because 
of the extremely low temperatures involved. Hence, numerical methods are necessary to aid our 
understanding of the structure of quantized vortices in different forms of turbulence, acting as guide 
for both experiments and theory. In this article, we shall discuss a widely used numerical model of QT, 
the vortex filament model (VFM).

\section{Vortex Filament Model}

In the VFM, vortices in the superfluid component are considered as line defects in which the phase 
changes by $2\pi$ when going around the core. In helium superfluids, the coherence length typically 
is much smaller than any other characteristic length scale. Therefore the VFM is a very suitable and 
convenient scheme to visualize the vortex dynamics in helium superfluids. Within the VFM the fluid 
velocity ${\bf v}_{\rm s}$ of the superfluid component is determined simply by the configuration of 
these quantized vortices and given by Biot--Savart law \cite{schwarz85}:
\begin{eqnarray}\label{e.bs}
{\bf v}_{\rm s}({\bf r},t) = \frac{\kappa}{4\pi}\int\frac{({\bf s}_1-{\bf r})\times d{\bf s}_1}
{\vert {\bf s}_1-{\bf r}\vert^3}\,.
\end{eqnarray}
Here, the line integration is along all the vortices and $\kappa = h/m$ is the circulation quantum. For
$^4$He $m = m_4$ is the bare mass of a helium atom (boson). In case $^3$He the condensation is made by Cooper 
pairs, and therefore $m=2m_3$. The Biot--Savart law expresses the Euler dynamics in integral form by assuming
a fluid of constant density \cite{saffman1992}.

Because the vortices are consider to be thin, the small mass of the vortex core can be neglected; therfore, 
at zero temperature, vortices move according to the local superfluid velocity. Numerically, the Biot--Savart 
integral is realized easily by having a sequence of points that describe the vortex. The singularity 
when trying to evaluate the integral at some vortex point, ${\bf s}$, can be solved by taking into 
account that the vortex core size, denoted by $a$, is finite \cite{schwarz85}:
\begin{equation}\label{e.bs2}
{\bf v}_{\rm s} = 
\frac{\kappa}{4\pi}\hat{\bf s}'\times {\bf s}'' \ln\left(\frac{2\sqrt{l_{+}l_{-}}}{e^{1/2}a}\right) + 
\frac{\kappa}{4\pi}\int^{'}\frac{({\bf s}_1-{\bf s})\times d{\bf s}_1}
{\vert {\bf s}_1-{\bf s}\vert^3}\,. 
\end{equation}
Here, $l_\pm$ are the lengths of the line segments connected to ${\bf s}$ after discretization, and
the remaining integral is over the other segments, not connected to ${\bf s}$. Terms $\hat{\bf s}'$ and 
${\bf s}''$, where the derivation is with respect to arc lengh, are (unit) tangent and normal 
at ${\bf s}$, respectively. The first (logarithmic) term on the right-hand side is the so-called 
local term, which typically gives the major contribution to ${\bf v}_{rm s}$. In the localized induction 
approximation (LIA), only this term is preserved (possibly adjusting the logarithmic factor). This 
is numerically convenient because the work needed per one time step will be proportional
to $N$ which is the number of points used to describe the vortex tangle. Including the nonlocal
term also will require $\mathcal{O}(N^2)$ operations. However, LIA is integrable, so in most cases, 
the inclusion of the nonlocal term is essential to break integrability. For example, under rotation, 
the correct vortex array is obtained only when the full Biot--Savart integral is used. 

At finite temperatures, the motion of quantized vortex is affected by mutual friction, which originates 
from scattering of quasiparticles from the vortex cores. Typically, the vortex motion may be described
by using temperature-dependent mutual friction parameters $\alpha$ and $\alpha'$, whose values are
well known \cite{bevan,Donnelly-Barenghi-alpha}. Then, the velocity of the vortex becomes \cite{HallVinen56,schwarz85}
\begin{equation}
{\bf v}_{\rm L}={\bf v}_{\rm s} +\alpha
\hat{\bf s}' \times ({\bf v}_{\rm n}-{\bf v}_{\rm s}) -\alpha'
\hat{\bf s}' \times [\hat{\bf s}' \times ({\bf v}_{\rm n}-{\bf v}_{\rm s})]\,.
\label{e.vl}
\end{equation}
This equation was derived by Hall and Vinen in the 1950s and was used by Schwarz few decades later, when
the first large scale computer simulations were conducted. This equation results when one balances the Magnus 
and drag forces acting on the filament. In general, the normal fluid velocity, ${\bf v}_{\rm n}$, should be 
solved self-consistently such that vortex motion is allowed to affect the normal component. This methodology 
may be applied, as in refs. \cite{Kivotides-science,Kivotides-coupled}; 
however, most studies in the literature have 
considered an imposed normal fluid velocity, ignoring any influence of the superfluid component on the normal 
component, which is more achievable numerically. Indeed, this is a reasonable approximation in $^3$He, where 
the normal component has a viscosity similar to that of olive oil and its motion is laminar. However, it is 
not appropriate in $^4$He, in which the normal component is extremely inviscid. Unfortunately, computational 
limits mean studies with full coupling have had limited scope up to now. For example, in ref. \cite{Kivotides-coupled}, 
the simulation was limited to an expanding cloud of turbulence and no steady state was reached. What is clear 
is that the next breed of numerical simulations should seek to follow this work and try to understand the 
dynamics of the fully coupled problem.

The presence of solid walls will alter the vortex motion, bacause the flow cannot go through the walls. 
For the viscous normal component, one typically uses the no-slip boundary condition, but for an ideal 
superfluid, the boundary condition is changed to no-flow through boundary, which implies that the vortex must 
meet the smooth wall perpendicularly. For plane boundaries, the boundary condition may be satisfied by using
image vortices, but with more general boundaries one has to solve the Laplace equation for the boundary
velocity field potential \cite{schwarz85}. This boundary velocity, plus any additional externally
induced velocities, generally must be included in ${\bf v}_{\rm s}$ and ${\bf v}_{\rm n}$ when determining the 
vortex motion using Eq.\,\eqref{e.vl}. If we are interested purely in homogeneous isotropic turbulence or 
flow far from the boundaries, then it is typical to work with periodic boundary conditions. These boundaries 
also may be approximated in the VFM by periodic wrapping; we duplicate the system on surrounding the 
computational domain with copies of itself: 26 in the case of a periodic cube, for example. The contribution 
of these duplicate filaments is then included in the Biot--Savart integral (Eq.\,\eqref{e.vl}).

\subsection{Reconnections} 
Vortex reconnections are essential in QT, allowing the system to be driven to a non-equilibrium steady 
state \cite{schwarz88}. They also change the topology of the tangle \cite{Poole-topo} and act to 
transfer energy from 3D hydrodynamic motion to 1D wave motion along the vortices \cite{Svistunov1995}. 
This phenomenon is important if we are to understand the decay of QT in the limit of zero temperature, 
which we discuss briefly towards the end of the article. Moreover, quantum vortex reconnections not 
only are important phenomena in quantum fluids, but also are relevant to our general understanding of 
fluid phenomena.

The VFM cannot handle vortex reconnections directly, because reconnections are forbidden by Euler dynamics. 
Therefore, an additional algorithm must be used, which changes the topology of two vortices when they
become close to each other, essentially a numerical ``cut and paste''. Several methods have been introduced to 
model a reconnection \cite{schwarz88, Konda2008, Tsubota2000,Baggaley-recon}. Importantly, a recent analysis 
\cite{Baggaley-recon} showed that all these algorithms produce very similar results, at least in case of 
counterflow turbulence. Microscopically, a single reconnection event was investigated by using the 
Gross--Pitaevskii model, which is applicable to Bose--Einstein condensates \cite{KoplikPRL1993,zuccher}. 
A recent numerical simulation of this microscopic model showed that the minimum separation between 
neighbouring vortices is time-asymmetric, like in classical fluids \cite{HD2011}. The VFM, on the other hand, 
results a more time-symmetric reconnection, in which the distance goes mainly as 
$d\propto \sqrt{\kappa|t-t_{\rm rec}|}$ where $t_{\rm rec}$ is the reconnection time \cite{zuccher,WA1994,PRB2013rec}. 
The prefactor, however, generally is somewhat larger after the reconnection event. This difference results 
from the characteristic curvatures, which are larger after a typical reconnection event \cite{BaggaleyCF,PRB2013rec}.

Interestingly, the results from the VFM are more compatible with experimental results \cite{Paolett-PYD}, 
with regard to the scaling of vortex reconnections. Although reconnections must be introduced ``by hand'' 
in the VFM, the model seems to capture the essential physics, at least at scales that currently can be 
probed experimentally.

\subsection{Tree-Code}
A potential drawback of the VFM is the computational time required to perform a simulation that captures 
the slowly evolving dynamics associated with the largest-scale motions. Although the LIA is computationally 
advantageous, several studies showed it to be unsuitable for studying fully developed QT \cite{Adachi2010}. 
However, as we have already alluded to, the inclusion of the nonlocal term in Eq.\,\eqref{e.bs2} 
means the scaling of the velocity computation is $\mathcal{O}(N^2)$. 
A similar problem arose in the field of computational astrophysics, in which calculations to compute the 
acceleration due to gravity also required $\mathcal{O}(N^2)$ operations. However, since the pioneering work 
of Barnes and Hut \cite{Barnes1986}, modern astrophysical and cosmological N-body simulations have made use 
of tree algorithms to enhance the efficiency of the simulation with a relatively small loss of 
accuracy \cite{Bertschinger1998}. The major advantage of these methods is the $\mathcal{O}(N\log(N))$ 
scaling that can be achieved. The essence of the method is to retain nonlocal effects but take advantage 
of the $r^{-2}$ scaling in Eq.\,\eqref{e.bs2}. Hence, the effect of distant vortices is reasonably small, 
and an average contribution may be used if it is computed in a systematic way. Several recent studies 
using the VFM made use of similar tree algorithms \cite{KivoRelax, Baggaleyfluc,BaggaleyCF} to achieve 
parameter regimes closer to those of actual experiments. It seems clear that tree methods, as in computational 
astrophysics, will become a standard addition to the VFM.

\subsection{Limitations}

The strongest limitation of the filament model is that it is based on the assumption that all the length scales 
considered are much larger than the vortex core size; therefore, the reconnections typically are made 
using the above cut and paste method. Numerical methods exist that model the vortex core structure and allow 
better handling of the reconnection process \cite{KivotidesEPL2003}. However, if the calculations are extended 
to core scales then all the slow large-scale phenomena associated, e.g. with vortex bundles, become numerically 
unreachable (even with a tree-code) because the time step (required for stability) typically scales as the square 
of the space resolution. Eulerian dynamics also prevent the generation of sound waves, which are allowed 
if the vortex dynamics in $^4$He is modeled by the Gross--Pitaevskii equation. Estimations by Vinen and Niemela 
\cite{VinenJLTP2002}, however, state that dissipation caused by phonon emission due to reconnections cannot 
fully explain the large dissipation observed in experiments at low temperatures. This estimation does not take 
into account the Kelvin wave (KW) cascade (see below) which is predicted to increase the sound emission. To conclude 
the above, the filament model still requires a physically justified method to model the dissipation at low 
temperatures, where the effect of mutual friction vanishes.

\section{Counterflow Turbulence}

The earliest experimental studies of QT were reported in a series of groundbreaking papers by Vinen in 
the 1950s \cite{Vinen-CF1,Vinen-CF2,Vinen-CF3,Vinen-CF4}. In these experiments turbulence was generated by 
applying a thermal counterflow, in which the normal and superfluid components flow in opposite directions. This 
is easily created by applying a thermal gradient, e.g., by heating the fluid at one end. The most common diagnostic 
to measure is the vortex line density, $L=\Lambda/V$,  where $\Lambda$ is the total length of the quantized 
vortices and $V$ is the volume of the system; from this, one can compute the typical separation between vortices, 
the inter vortex spacing $\ell=1/\sqrt{L}$. This can readily be measured experimentally using second 
sound \cite{DonnellyBook}, and higher harmonics can probe the structure of the tangle. Numerical simulations 
have played a crucial role in visualizing the structure of counterflow turbulence and probing the nature of 
this unique form of turbulence; indeed it has no classical analog. Some of the very earliest studies using 
the VFM were performed by Schwarz \cite{schwarz88}; however, computational limitations forced him to 
perform an unphysical vortex mixing procedure. A more recent study by Adachi \etal \cite{Adachi2010} 
made use of modern computational power and studied the dependence of the steady-state vortex line density on 
the heat flux of the counterflow. Within the parameter range of the study, there was a good agreement with 
experimental results, vindicating the use of the VFM for counterflow turbulence. 

A particularly striking example in which the power to visualize QT helped answer an 
apparent puzzle is the anomalous decay of counterflow turbulence. Experimentally, it was observed that after 
switching off the heater, the density of quantized vortices decreased in time, as was expected. In the 
early stages of the decay, the process was very slow, and it even was observed that the vortex line density 
might increase, after the drive was turned off. An explanation was provided in ref. \cite{Depolarization}, in which 
the authors showed, using the VFM, that the tangle created by counterfow is strongly polarized. However, after 
they switched off the drive, the vortex lines depolarize. As experimental measurements are based on second sound, 
this depolarisation creates an apparent increase in the measured vortex line density, which is purely an 
artifact of the measurement process, a truly beautiful result.

A more recent study by Baggaley \etal \cite{BaggaleyCF} probed the structure of the superfluid vortices 
in thermal counterflow by convolving the tangle with Gaussian kernel to identify any structures in the tangle. 
In contrast to counterflow, QT generated by more conventional methods, such as mechanical 
stirring, exhibits the famous Kolmogorov scaling, $E(k) \sim k^{-5/3}$ \cite{Tabeling}. It is 
expected \cite{L'vov2007}, and has been observed numerically \cite{BaggaleyLaurie}, that this $k^{-5/3}$ 
scaling is associated with vortex bundling, which may mediate energy transfer in the inertial range. 
The study of Baggaley \etal \cite{BaggaleyCF} showed that this is not the case in counterflow, with the 
tangle being relatively featureless and consisting of random closed loops, leading to a featureless 
energy spectrum without motion of many scales, which is the hallmark of classical turbulence.

\begin{figure}[!ht]
\centerline{
\includegraphics[width=8.4cm]{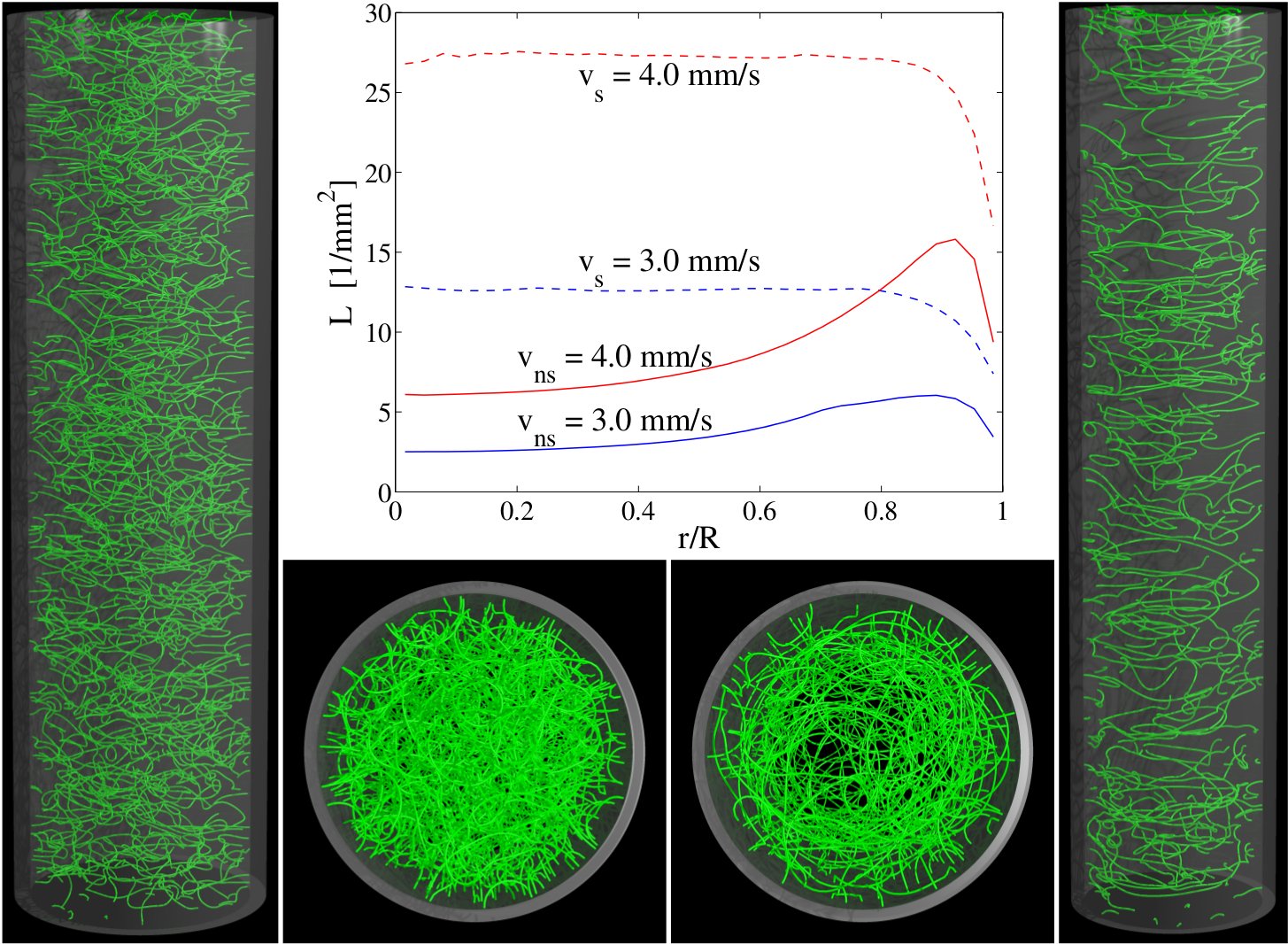} 
}
\caption{Counterflow in a cylinderical pipe of radius $R$ = 1\,mm at $T$ = 1.9\,K. For vortex structures on 
the left, the counterflow is caused by pure superflow $v_{\rm s}$ = 4.00\,mm/s, and on the right, both components are 
involved in gerating the same counterflow ($v_{\rm s}$ = 1.68\,mm/s and $v_{\rm n}$ = -2.32\,mm/s) such that the total 
massflow is zero. For the normal component, a parabolic profile is used. 
The middle part of the figure illustrates the vortex line density profile inside the pipe for counterflows of 
3.00 mm/s and 4.00 mm/s, using both pure superflow (dashed lines) and thermal counterflow (solid lines), 
when averaged over a wide time window in the steady state.  
}\label{fig_cflow}
\end{figure}

Visualization of the counterflow turbulence and detailed analysis of the nature of the flow the vortices 
induce also have direct consequences for analytic theories of QT. For example, Nemirovskii 
\etal \cite{Nemirovskii-Tsubota-Araki-2002} considered the energy spectrum of the velocity field induced by 
a random set of quantized vortex rings. The analytic spectrum they predicted is quantitatively similar to 
the energy spectrum obtained numerically by Baggaley \etal \cite{BaggaleyCF}.

There still are several open and important questions related to the problem of counterflow turbulence. 
In particular, most numerical simulations have considered the flow away from boundaries to justify the use 
of periodic boundary conditions. However, this approximation ignores a large amount of important and 
interesting physics. In the remainder of this section, we consider both counterflow and superflow along a pipe, 
in which the effects of boundaries are included. Experimentally, pure superflow, in which the normal fluid is held 
static, is possible using superleaks at the ends of the pipe. Although initially this may seem purely a 
Galilean transform of the problem of counterflow, results \cite{Chagovets} indicate differences between the 
two states. However, one should note that for normal fluid one should have a no-slip boundary condition at 
the boundaries. Therefore, even with laminar (parabolic) normal component, the counterflow profile is not 
fully flat (unlike pure superflow) when both components are involved in generating the counterflow. Using 
the VFM, we observe that for thermal counterflow, in which the velocity of the normal component is nonzero, 
a larger average counterflow is needed to achieve a similar vortex line density than when using pure superflow.
It is possible that a smaller value of the velocity difference between the two components ($\bv_{\rm n}-\bv_{\rm s}$), 
near the boundaries, may explain this result. Turbulence is caused by the vortex instability at the 
boundaries \cite{deGraafJLTP2008}, and the level of turbulence that can be supported depends on the 
counterflow velocity at the boundaries. These results are illustrated in Fig.\,1 in which the simulations of 
counterflow and pure superflow in a pipe with a fixed average flow rate are presented. Future work, 
following ref. \cite{BaggaleyFP2013}, still should be undertaken to further investigate the role of 
normal fluid turbulence on the observed vortex configuration and vortex line density \cite{Tough}.

\section{Two-Fluid Turbulence}

Whereas thermal counterflow is a unique form of turbulence, possible only in  quantum fluids, one of the 
main motivations behind the study of QT is in the so-called semiclassical regime, in which the statistical 
properties of the turbulence show tantalising similarities to normal viscous fluids. In particular, this 
regime has been the focus of several experimental studies \cite{Tabeling,Salort2011} in which the classical 
Kolmogorov energy spectrum and higher order statistical measures, such as structure functions, show agreement 
with classical studies. This result also was reproduced by Araki \etal \cite{Araki}, who use the VFM at 0\,K, 
to study the evolution of quantized vortices arranged as the classical Taylor--Green vortex.

In addition, the VFM has played an important role in allowing us to visualize the structure of the quantized 
vortices under the influence of a turbulent normal fluid. Morris \etal \cite{Koplik-Morris} performed a 
particularly influential study in which they coupled a full numerical simulation of the Navier--Stokes 
equation to the VFM. They observed a locking between vortices in the superfluid component and intense vortical regions 
in the normal component. This finding built upon an earlier study by Kivotides \cite{KivotidesPRL2006} in which 
a similar result was obtained, but for a frozen normal fluid velocity field, generated by a turbulent tangle 
of classical vortex filaments.

In a more recent paper, Kivotides \cite{KivoRelax} considered the effect of coherent superfluid vortex bundles 
on an initially stationary normal fluid.  Computations were performed using the VFM coupled to the Navier--Stokes 
equation, with mutual friction accounted for as a forcing term in the Navier--Stokes equation. The author showed 
that the induced normal-fluid vorticity acquired a morphology similar to that of the structures in the superfluid 
fraction, and argued that the dynamics of fully developed, two-fluid turbulence was depended on interactions of 
coherent vortical structures in the two components.

Indeed, in classical turbulence, these nonlinear structures, vortical ``worms'', appear to play a crucial role 
in the dynamics of the inertial range \cite{Farge}. In a more recent study, Baggaley \etal \cite{BaggaleyLaurie} 
developed a procedure to decompose the vortex tangle into a coherent ``bundled'' component and a random component. 
Algorithmically, this was achieved by convolving the vortex tangle with a cubic spline:
\begin{equation}
\bom(\bs_i)=\kappa \sum_{j=1}^N \bs'_j W(r_{ij},h) ds_j,
\label{eq:smooth}
\end{equation}

\noindent
where $r_{ij}=|\bs_i-\bs_j|$, $ds_j=|\bs_{j+1}-\bs_j|$,
$W(r,h)=g(r/h)/(\pi h^3),$ 
$h$ is a characteristic length scale, and
\begin{equation}
g(q) =
\begin{cases}
1 - \frac{3}{2}q^{2} + \frac{3}{4}q^{3} , & 0 \leq q < 1 ; \\
\frac{1}{4}\left( 2 - q\right)^{3}, & 1 \leq q < 2 ;\\
0, & q \geq 2.
\end{cases}
\label{eq:spline}
\end{equation}
It is appropriate to take $h$ equal to the intervortex spacing, effectively smearing the quantized 
vorticity to create a continuous vector field in space. Thus, at any point, an effective ``vorticity'' may be 
defined; in particular, vortex points with a high vorticity (above a threshold level based on the root-mean-squared 
vorticity) were categorized as part of the coherent component. Analysis of these two components showed 
that it is the vortex bundles which create the inertial range of the turbulence and the random component 
simply is advected in the manner of a passive tracer. 
 
In the next section, we consider quantum fluids under rotation, in which we apply this smoothing algorithm 
to identify previously unidentified transient coherent structures that appear in the system.

\section{Coherent Structures under Rotation}

Rotating fluids are ubiquitous in the universe, so the study of classical fluids under rotation forms a vast 
topic in its own right. Within the field of viscous fluids several different flow profiles have been observed, 
depending on external conditions etc. In helium superfluids rotation, has been used actively to investigate 
vortex dynamics. The steady state under constant rotation typically is a vortex array that mimics the 
normal fluid profile. However, before this steady state is reached, turbulence may appear, especially 
at low temperatures when the mutual friction is low \cite{nature2003,EltsovPLTP16,EltsovJLTP2010}. 
The onset and initialization of turbulence has been attributed to the instability that originates from 
interaction of the vortex with the container walls \cite{FinnePRL2006prec,deGraafJLTP2008,EltsovJLTP2010}.

\subsection{Vortex Front}

Recently, perhaps the most investigated coherent structure that appears under rotation has been a propagating 
turbulent vortex front, which separates a vortex-free region from a twisted vortex cluster behind the front
\cite{twist2006,front2007,front2011}. The front may be observed in superfluid $^3$He-B, because the critical
velocity for vortex nucleation due to surface roughness can adjusted  to be large enough 
so that a vortex free rotation (the so-called Landau state) can be sustained even at relatively large 
rotation velocities. Now, if vorticity is introduced--for example, by using the Kelvin--Helmholtz instability of 
the A-B phase boundary \cite{BlaauwgeersPRL2002}--a front is generated easily. The propagation velocity of the 
front is proportional to the dissipation. At the lowest observable temperatures ($T \sim 0.15T_{\rm c}$) the coupling 
with the normal fluid almost vanishes, but the energy dissipation still is observed to be finite, orders of magnitude 
larger than one would obtain from the laminar prediction \cite{front2007}. However, the dissipation of angular 
momentum remains weak, which is seen from the rotation velocity of the vortex array behind the front. At lowest 
temperatures, this rotation velocity drops much below the rotation velocity of the cylindrical cell \cite{front2011}. 
In a way, the vortex motion decouples from the external reference frame of the rotating cylinder. All these 
features also are observed in vortex filament simulations \cite{front2007,front2011}. Most recently, the VFM helped 
to develop a simple model, which explains the observed behavior \cite{front2013}.

\subsection{Vortex Bundles During Spin-Up}

\begin{figure}[!t]
\centerline{
\includegraphics[width=8.4cm]{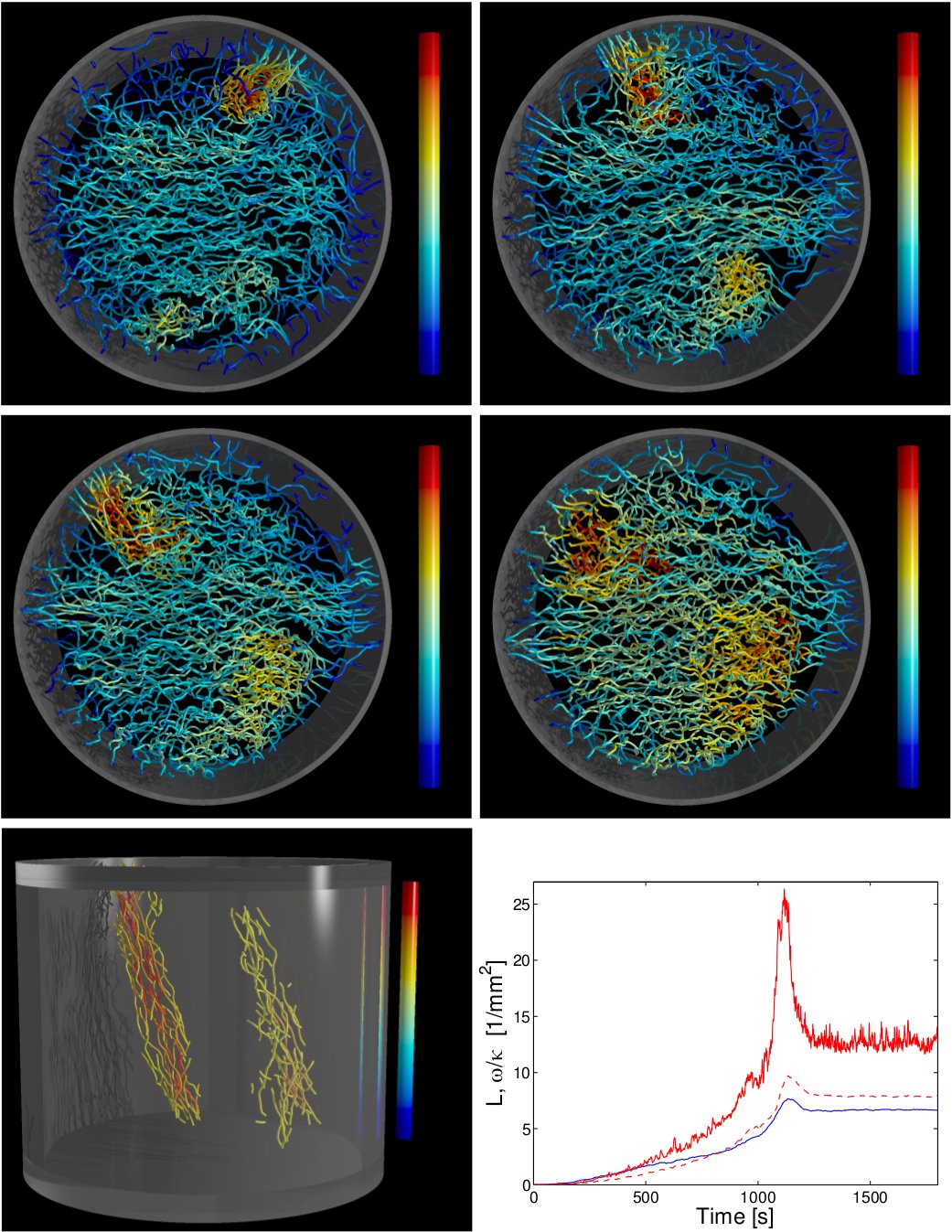}
}
\caption{
Coherent structures appearing in a tilted cylinder during the spin-up of the superfluid component.
Superfluid $^3$He-B at $T$ = 0 with $R$ = 3 mm, $L$ = 5 mm, $\Omega$ = 0.25 rad/s, and a tilt angle of 30$^\circ$.
The configurations are shown at times $t$ = 1100 s (Top Left), 1130 s (Top Right), 1140 s 
(Middle Left), and 1150 s (Middle Right), and the color coding illustrates the relative amplitude 
of the smoothed vorticity. (Bottom Left) The coherent structures at $t$ = 1140 s, 
where only the coherent part with $\omega > 1.4\omega_{\rm rms}$ is plotted. (Bottom Right) The temporal 
evolution of the vortex line density, $L$ (solid blue line), together with the 
rms (dashed red line) and maximum vorticity (solid red line).  
}\label{f.cohspinup}
\end{figure}

The structure of the vortex front is a rather natural outcome, bacause the equilibrium state is a vortex array which 
tries to mimic the solid-body rotation of the normal component. Here, we have identified a coherent 
vortex structure by applying cubic spline smoothing, Eq.\,\eqref{eq:smooth}, on the vortex structures that 
appear during the spin-up (by suddenly increasing the rotation velocity) of the superfluid component. Our geometry
is a cylinder that is strongly tilted with respect to the rotation axis \cite{HanninenJLTP2009}. Initially 
we have only a single vortex present, which expands because of the applied flow that is the result of rotation and 
is nonzero even at zero temperature because of the tilt of the cylinder. As time passes, a polarized vortex tangle  
develops, which eventually approximates solid-body rotation. Figure 2 
illustrates what happens slightly before the configuration reaches the steady state. Two localized vortex 
structures (vortex bundles) appear on the opposite sides of the cylinder. In simulations, they are observed at low 
temperatures with small mutual friction and they appear during the ``overshoot'' period and quickly merge to the 
background vorticity. It also is interesting to note that the steady state that approximately mimics the solid-body
rotation is reached even at zero temperature, at which the mutual friction coefficients are set to zero. 
Naturally, there is some small numerical dissipation. A somewhat more peculiar feature of these simulations 
is that the steady state is not fully static, even at somewhat higher temperatures. This state might appear
because the boundary induced velocity in these simulations is solved only approximately (by using image vortices)
\cite{HanninenJLTP2009}.  Alternatively, the simulations perhaps are stuck in some other local energy minimum, 
which is not the true minimum. However, these two vortex bundles are just one more example of how superfluid 
can mimic classical fluids at length scales larger than the intervortex distance by forming coherent structures,
even at very low temperatures at which the normal component is vanishingly small.

\subsection{Spin-Down}

The decay of quantized vortices at low temperatures after a sudden stop of rotation (spin-down) has 
been analyzed in several experiments, in both superfluid $^3$He-B and $^4$He-II. The $^3$He-B 
experiments conducted in a cylindrical container, show a laminar-type decay in which the vorticity typically 
decays as $1/t$ \cite{spindown2010,EltsovJLTP2010}. In contrast, the experiments with $^4$He-II, 
using a cubical container, show a turbulent decay in which vorticity 
decreases faster, proportional to $t^{-3/2}$, and is preceded by a strong overshoot, just after the rotation 
stops \cite{WalmsleyPRL2007,EltsovPLTP16}. Although the stronger pinning in $^4$He-II may favor turbulence 
over laminar behavior, the recent simulation that used smooth walls, in which pinning was neglected, showed 
that geometry has a strong effect on decay behavior at low temperatures. Simulations conducted 
in a sphere, or in a cylinder in which the cylinder axis is close to the initial rotation axis, show a 
laminar decay in which the vortices remain highly polarized. In contrast, calculations performed in a 
cubical geometry, or in cases in which the tilt angle for the cylinder is large, indicate turbulent decay
\cite{spindown2010,EltsovJLTP2010}.

In cylindrically symmetric containers, the decay of vorticity is observed to occur in a laminar fashion, 
which may be explained by using the Euler equation for invisid and incompressible flow in uniform rotation. 
If the initial vorticity corresponds to an equilibrium state given by the initial rotation, $\Omega_0$, and
if the rotation is set to rest, then the solution for radial part of the 2D motion of the 
vorticity $\Omega_{\rm s}(t)$ is given by $\Omega_{\rm s}s = \Omega_0/(1+t/\tau)$ \cite{SoninRMP1987,EltsovJLTP2010}. 
The decay time is given simply by the mutual friction as $\tau = 1/(2\alpha\Omega_0)$. This appropriately 
models the vorticity in the bulk, in which the polarization is near 100\% and the coarse-grained 
vorticity is spatially homogeneous. In simulations, the dissipation of vorticity (vortex line length) 
occurs within a thin boundary layer whose thickness increases as temperature (mutual friction dissipation) 
decreases \cite{EltsovJLTP2010}. What happens in the zero temperature limit remains a somewhat open 
question. Presumably, the thickness of the boundary layer increases as $T \rightarrow 0$ so that the 
laminar decay becomes impossible with vanishing mutual friction. 

In cubical containers, or if the cylindrical symmetry is broken, \emph{e.g.}, by strongly tilting the
cylinder, the decay shows turbulent behavior, even if the polarization remains nonzero (reflecting the
long-surviving vortex array). After an initial overshoot, the decay is faster than in the laminar case. 
Reconnections here are distributed more evenly in the bulk, indicating turbulence in the whole volume.
In addition to the vortex array, the most visible indication of coherent structures is the helical type 
distortions of this array. This phenomenon is illustrated in Fig.\,3, 
in which we have applied the above vortex smoothing procces, Eq.~\eqref{eq:smooth}, 
on recent spin-down simulations conducted in a cube. These coherent oscillations of 
the vortex array appear shortly after the rotation is stopped and might be related to the
lowest inertial wave resonances. Because the smoothed vorticity resulting from the vortex array is rather 
uniform and the fluctuations from this level are quite small, the numerical identification of 
additional structures, if they exist, becomes difficult. However, the faster decay and the apparent 
absence of different size coherent structures, typical for Kolmogorov turbulence, might indicate 
that the decay of type $t^{-3/2}$ is more general than expected.

\begin{figure}[!b]
\begin{center}
\includegraphics[width=8.4cm]{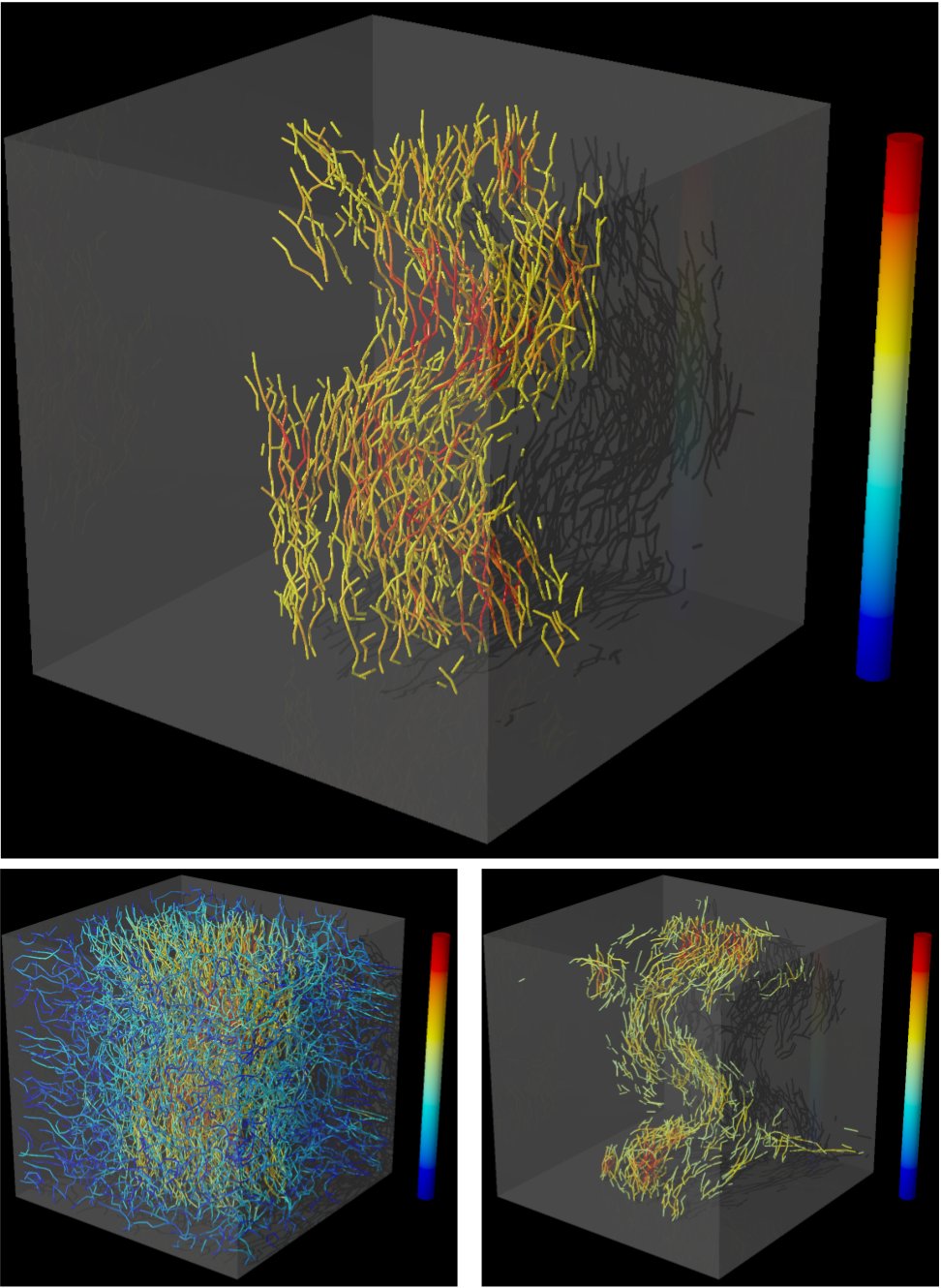}
\end{center}
\caption{
Decay of vorticity in superfluid $^3$He-B after a sudden stop of rotation from $\Omega_0$ = 0.5 rad/s. 
The initial state inside the cube (side 6 mm) was a steady-state vortex array, with a small tilt to break 
the symmetry. (Upper) The coherent vortex part ($\omega > 1.4\omega_{\rm rms}$) at 
$T$ = 0.20$T_{\rm c}$ with $t$ = 87 s. (Lower Left) The full vortex configuration 
with $\omega_{\rm rms}$ = 9.23 s$^{-1}$ and $\omega_{\rm max}$ = 21.05 s$^{-1}$. (Lower Right)
The coherent part but at slightly higher temperature, $T$ = 0.22$T_{\rm c}$, 
and $t$ = 64 s. Here $\omega_{\rm rms}$ = 10.90 s$^{-1}$ and $\omega_{\rm max}$ = 29.23 s$^{-1}$. Coloring of 
the lines shows the smoothed vorticity, normalized by the maximum smoothed vorticity.
}\label{f:spindown}
\end{figure}

\begin{figure}[!ht]
\begin{center}
\includegraphics[width=8.4cm]{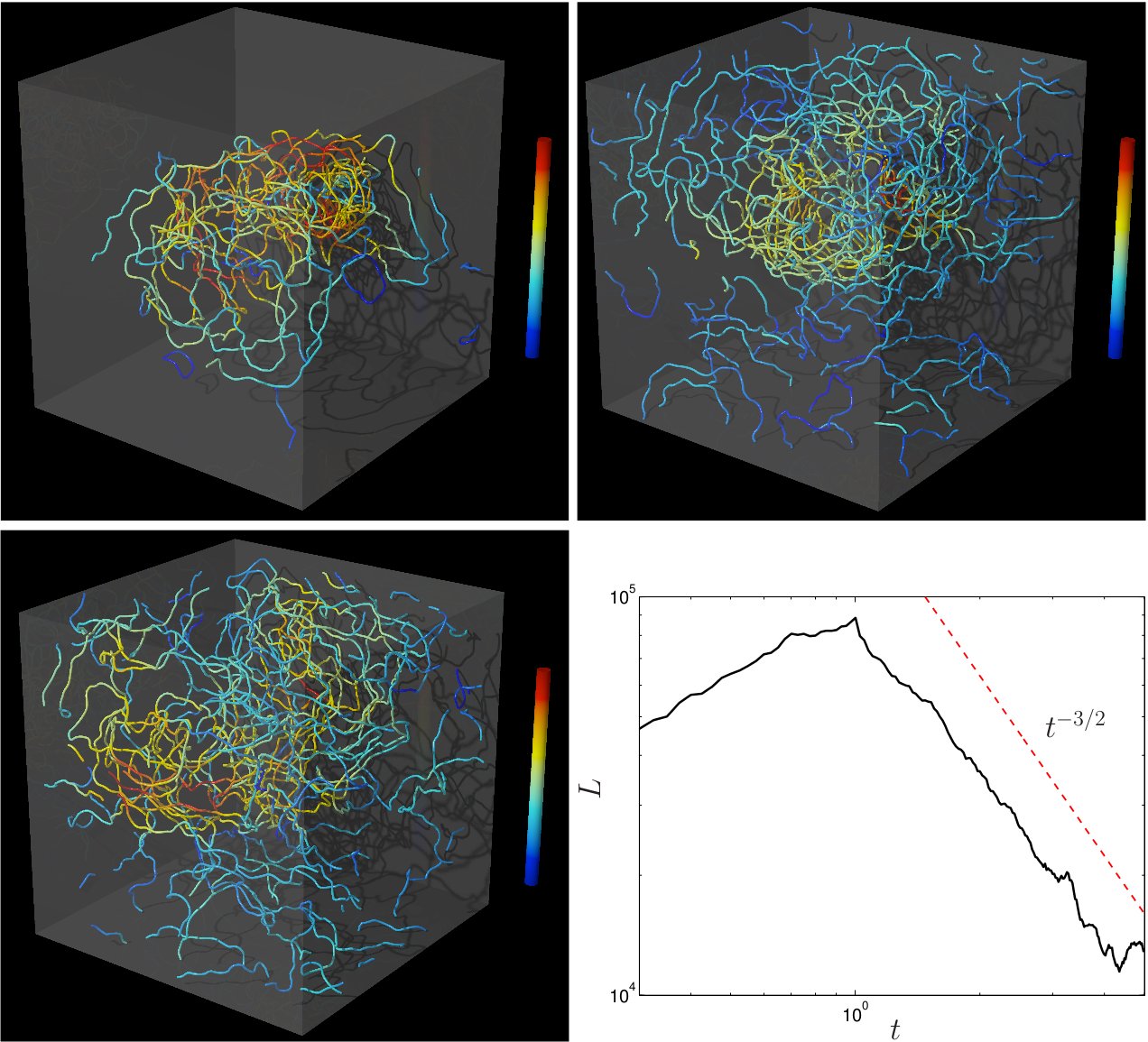} 
\end{center}
\caption{
Decay of semiclassical turbulence in superfluid $^4$He. The vortex configurations are plotted at 
$t$ = 0.1\,s (Upper Left), $t$ = 0.75\,s (Upper Right) and $t$ = 1.1\,s (Lower Left). 
The coloring of the lines shows the smoothed vorticity, normalized by the maximum smoothed vorticity 
($\omega_{\rm max}$ = 16.31, 37.7 and 29.3 s$^{-1}$ respectively.) The shaded box shows the periodic domain of 
the simulation, a cube of side 0.03\,cm. The corresponding vortex line densities are 4.1$\times$10$^4$, 7.8$\times$10$^4$ 
and  7.5$\times$10$^4$\,cm$^{-2}$. (Lower Right) The evolution of the vortex line density and the 
semiclassical scaling. 
}\label{f:semiclassic}
\end{figure}

\section{The Decay of a Random Tangle}

Of course, the study of turbulent decay is not limited to rotating cases; indeed, the decay of homogeneous 
isotropic turbulence is an important field of research. Here, we focus on the decay of QT in the limit of 
zero temperature. Towards the end of the article, we focus on the physical mechanisms of energy 
dissipation; here we limit ourself to the scaling of the decay, as this also is readily measured 
experimentally. Experiments in helium have revealed two distinct regimes of decay of a random tangle of 
quantized vortices by monitoring the vortex line density $L$ in time. These are the so-called ultraquantum 
decay, characterized by $L \sim t^{-1}$, and semiclassical, $L\sim t^{-3/2}$, regimes. Perhaps the most 
striking example of these two regimes came from a study by Walmsley and Golov \cite{Walmsley2008}. By injecting 
negative ions into superfluid $^4$He, in the zero temperature limit, they observed the two regimes of 
turbulence decay. The negative ions (electron bubbles) generated vortex rings, which subsequently interacted, 
forming a turbulent vortex tangle. After switching off the ion injection the turbulence then would decay. 
If the injection time was short, the ultraquantum regime was observed, whereas for a longer injection 
time semiclassical behaviour was apparent. 

Walmsley and Golov argued that the second regime is associated with the classical Kolmogorov spectrum 
at low wave numbers, whereas the ultraquantum regime is a result of the decay of an unstructured tangle 
with no dominant large scale flow. Both regimes also were observed in $^3$He-B by Bradley \etal 
\cite{Bradley2006}, who forced turbulence with a vibrating grid.

In a numerical study using the VFM, Fujiyama \etal \cite{Fujiyama2010} showed some evidence that a 
tangle generated by loop injection might exhibit semiclassical behaviour and decay such as $L\sim t^{-3/2}$. 
A more comprehensive study by Baggaley \etal \cite{Baggaley-decay} drew inspiration from the experiment 
of Walmsley and Golov and considered both short and long injection time. This study reproduced both 
the ultraquantum and the semiclassical regimes and by examining both the curvature of the filaments 
and the superfluid energy spectrum, confirmed the hypothesis of Walmsley and Golov. In the semiclassical 
regime, the initial energy distribution was shifted to large scales and a Kolmogorov spectrum was formed. 
If we now revisit the structure of the tangle in the semiclassical simulation and perform the 
convolution with a cubic spline, Eq.\,\eqref{eq:spline}, then we can see coherent bundling of the vortices, 
potentially as a result of the strong anisotropy in the loop injection, (Fig.\,4). 
In the ultraquantum case, as the injection time is short, the spectrum decays without this energy transfer 
and very little energy is in the large scale motions.

\section{Coherent Structures due to Normal Fluid Shear}

Having identified the presence of coherent structures in various systems, it is natural to turn our 
attention to the generation of coherent structures in QT. Previous studies have focused on the role of 
intense, localized, vortical structures in the normal fluid component \cite{Koplik-Morris,KivotidesPRL2006}. 
The origin of these classical vortex structures is often attributed to the roll-up of vortex sheets by 
the Kelvin--Helmholtz instability \cite{VincentJFM94}. Here, we demonstrate that in QT such a 
mechanism is viable and that simply the presence of shear in the normal component is enough to 
lead to the generation of coherent structures. We consider a numerical simulation using the VFM of QT 
driven by an imposed shear flow in the normal component at $T$ = 1.9\,K. The normal fluid profile is given 
by $\bv_{\rm n}= A(\sin(2\pi y/D),0,0)$, where $D$ = 0.1\,cm is the size of the (periodic) domain and $A=2\,{\rm cm/s}$. 
The act of the shear in the normal component is to concentrate vortices into areas of low velocity, where 
they form vortex sheets. In these sheets, vortices lie approximately parallel; therefore, large vortex line 
densities are created as the dissipative effect of reconnections is small. Once the vortex density in 
these areas becomes large enough, the sheets visibly buckle and begin the roll-up process, 
as may be seen in Fig.\,5. This effect might be important in a several of scenarios, particularly in 
the onset of normal fluid turbulence in counterflow.

\begin{figure}[!th]
\begin{center}
\includegraphics[width=8.4cm]{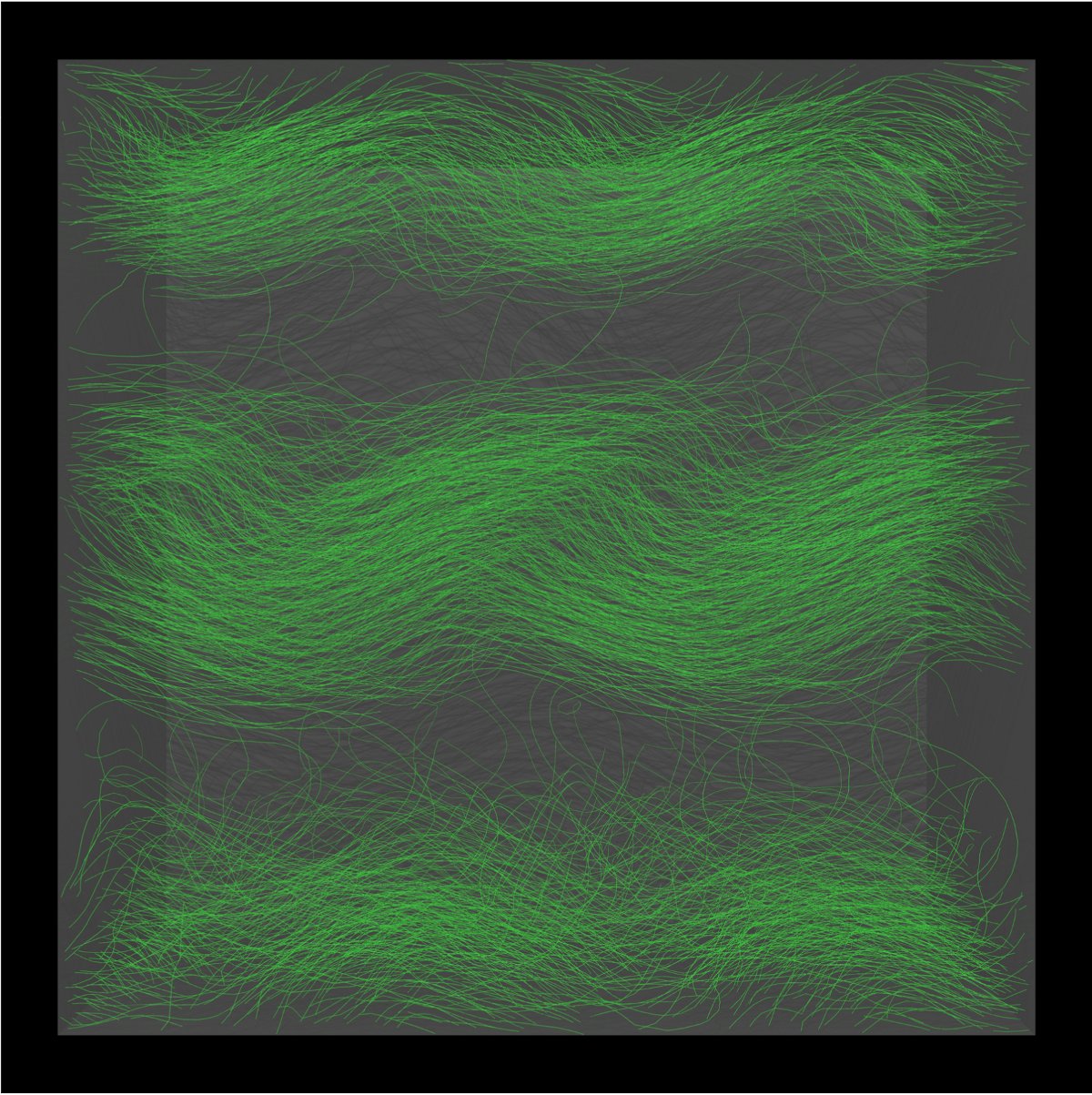}
\end{center}
\caption{
Sheets of quantized vortices beginning to roll up as a result of the Kelvin--Helmholtz instability. 
The vortex sheets are created by an imposed shear in the normal component.}\label{f:shear}
\end{figure}

\section{Future Challenges}

In addition to topics covered here, the VFM also has shown that it may be a useful tool in interpreting 
the experiments in which the motion of tracer particles is used to visualize the quantum turbulence, or in 
investigating the properties of the vortex tangle generated by oscillating structures. In the future, coupled
dynamics with normal fluid will be essential in understanding the dynamics of the normal component. Currently,
the dissipation in the zero temperature limit is perhaps the most interesting topic in the field of QT. 
In the following sections, we concentrate on two topics strongly related to this, namely the KW cascade and 
the possible bottleneck appearing at scales of the order the average intervortex distance.

\subsection{Kelvin Wave Cascade}

{\mbox\,At} relatively large temperatures (greater than 1K in $^4$He), 
kinetic energy contained in the superfluid component is transferred by mutual 
friction to the normal fluid, and subsequently into heat via viscous heating. 
A constant supply of energy (continuous stirring, for example) is needed to maintain the 
turbulence. At very low temperatures, the normal fluid is negligible, but despite the 
absence of mutual friction, the turbulence still decays \cite{Bradley2006,Walmsley2008}.
The KW cascade \cite{Svistunov1995} perhaps is the most important mechanism proposed 
to explain this surprising effect.

KWs are a classical phenomena \cite{LordKelvin}, rotating sinusoidal or helical 
perturbations of the core of a concentrated vortex filament. The KW cascade is the process in which 
the nonlinear interaction of KWs creates higher frequency modes. At very high frequencies 
(at which the wavelength is atomic scale), sound is efficiently radiated away (phonon emission). 
Hence, in contrast to classical turbulence, in which the energy sink is viscous, in QT the energy 
sink is acoustic. Crucially, KWs are generated easily in QT, in which vortex reconnections 
typically create a high-curvature cusp \cite{Kivotides-cascade}, which acts as a mechanism to 
transfer energy from 3D hydrodynamic turbulence, to 1D wave turbulence along the vortex filaments.

Currently two regimes are believed to exist in the KW cascade, one corresponding to large-amplitude 
waves and the another to a low-amplitude, weakly nonlinear regime, in which the theory of 
wave turbulence can be applied \cite{Nazarenko2011}. It is in this weakly nonlinear regime where 
the VFM may play a crucial role in distinguishing several proposed theories 
\cite{Kozik-Svistunov-PRL-2004,LN2010,SoninPRB2012}. The key prediction of each theory is in the 
spectrum of the kelvon occupation numbers, each of which gives different power-law scalings, 
$n_k \sim k^{-\alpha}$. In particular Kozik and Svistunov \cite{Kozik-Svistunov-PRL-2004} 
proposed $\alpha=17/5$, but L'vov \etal \cite{LN2010} claimed this spectrum was invalid because of the 
assumption of locality of interactions and proposed a nonlocal theory that predicted $\alpha=11/3$. 
Although it is not simply the spectrum of $n_k$ that distinguishes these two theories, it
perhaps is the easiest statistic to compute with the VFM, as $n_k$ is related to the KW 
amplitudes. However, the fact that these exponents are so similar clearly presents a huge 
computational challenge if one is to provide strong evidence for either theory. 
Few attampts have been made in determining the exponent $\alpha$ 
\cite{Vinen-cascade,Kozik-Svistunov-PRL-2005,BaggaleyPRB2011spectrum}, but we would argue 
that no convincing evidence for either theory has been demonstrated yet. One reason 
is the difficulty in identifying a KW on a curved vortex \cite{HietalaJLTP2012}.


\subsection{Bottleneck?} Although much attention has been focused on the KW cascade, perhaps of more importance, 
particularly in experimental interpretation, is how the 1D KW cascade matches the 3D hydrodynamic 
energy spectrum. Once again rival theories have been proposed by Kozik \& Svistunov and 
by L'vov \etal. To summarize the situation briefly, L'vov \etal predict a bottleneck in energy, 
which is required for continuity in the energy flux at the crossover scale; therefore, one should expect 
an increase in the vortex line density at scales on the order of the intervortex spacing, $\ell$. 
This is countered by Kozik \& Svistunov \cite{Kozik-Svistunov-PRB-2008}, who argue for several 
different reconnection regimes between the Kolmogorov and KW spectra that do not create 
such a bottleneck. However, another view has been provided by Sonin \cite{SoninPRB2012}, who argues that 
the bottleneck might be totally absent. Again, this is an open and important question that has yet to 
be studied in detail using the VFM. The large range of scales involved means that new numerical approaches, 
such as the tree-code discussed earlier, will be vital for any progress to be made.

\section{Conclusions}

To summarize, we hope the reader will agree that the VFM has proven to be a valuable tool in the 
study of superfluid turbulence and quantized vortex dynamics. Here, we have illustrated that the quantized 
vortices can form coherent structure, even at low temperatures at which the fraction of normal component is small, 
and give QT a classical nature. The formation of coherent structures also is shown to appear easily in 
the presence of the normal fluid shear, resulting in the roll-up of vortex sheets due to classical 
Kelvin--Helmholtz instability. Of course, despite some of the success stories we have described here, there 
is still much work to be done. Much attention in the literature has focused on the KW cascade, 
and perhaps rightly so. However, other decay mechanisms, such as loop emission due to vortex reconnections 
\cite{KBL2011}, warrant further investigation. Indeed, they may play an important role in the decay of 
the unstructured ``Vinen tangle.''

The field itself will of course will be driven by experimental studies, and it is an exciting time with many 
new investigations planned in the near future. Nevertheless, the role of the VFM in helping to analyse and 
interpret this experimental data and test, refine, and motivate analytic theories remains important.

In addition, superfluid turbulence is not just found in the laboratory. There are important astrophysical 
applications. Current theory strongly suggests that the outer core of neutron stars consists of neutrons 
in a superfluid state. Because of incredibly rapid rotation, we would expect this superfluid to be threaded by 
quantized vortices pinned to the solid outer crust. Interesting phenomena, such as rapid changes in the 
rotation rate, are observed and thought to be related to the behaviour of the quantized vortices 
\cite{Anderson2012}. Such a system might reasonably be modelled with the VFM, and it remains an interesting 
problem awaiting such an investigation.





\begin{acknowledgments}
This work is supported by the European Union 7th Framework Programme (FP7/2007-2013, grant 228464 Microkelvin).
R.H. acknowledges financial support from the Academy of Finland. 
R.H. would also like to thank N. Hietala for useful discussions and 
CSC - IT Center for Science, Ltd., for the allocation of computational resources. 


\end{acknowledgments}





\end{article}








\end{document}